\documentstyle[aps,prl,epsf]{revtex}
\begin{document}
\draft
\title{Characterization of the Emergence of Order in an Oscillated Granular Layer}
       
\author{Daniel I. Goldman,$^{1}$ Gemunu H. Gunaratne,$^{2,3}$
        M. D. Shattuck,$^{1}$ Donald J.  Kouri,$^{2,4}$ David
        K. Hoffman,$^{5}$ D. S. Zhang,$^{4}$ and Harry
        L. Swinney$^{1}$}
        
\address{$^{1}$ Center for Nonlinear Dynamics,
                The University of Texas at Austin,
                Austin, TX 78712}
\address{$^{2}$ The Department of Physics,
                University of Houston,
                Houston, TX 77204}
\address{$^{3}$ The Institute of Fundamental Studies,
                Kandy, Sri Lanka}
\address{$^{4}$ The Department of Chemistry,
                University of Houston,
                Houston, TX 77204}
\address{$^{5}$ Department of Chemistry and Ames Laboratory,
                Iowa State University,
                Ames, IO 50011}

\nobreak
\maketitle
\begin{abstract}
The formation of textured patterns has been predicted to occur in two
stages. The first is an early time, domain-forming stage with dynamics
characterized by a disorder function $\bar\delta (\beta) \sim
t^{-\sigma_{E}}$, with $\sigma_{E} = \frac{1}{2}\beta$; this decay is
universal. Coarsening of domains occurs in the second stage, in which
$\bar\delta (\beta) \sim t^{-\sigma_{L}}$, where $\sigma_{L}$ is a
nonlinear function of $\beta$ whose form is system and model
dependent. Our experiments on a vertically oscillated granular layer
are in accord with theory, yielding $\sigma_{E}\approx 0.5\beta$, and
$\sigma_{L}$ a nonlinear function of $\beta$.
\end{abstract}  

The formation of textured spatial patterns has been studied in
laboratory experiments~\cite{expt} and model
systems~\cite{croAhoh}. Several quantities have been used to describe
the development of patterns represented by a scalar field $v({\bf x})$
with a typical wavevector $k_0$, including the structure
factor~\cite{eldAvin,croAmei,chrAbra,schAall}, density of topological
defects~\cite{houAsas}, and a recently introduced family of
characterizations called the disorder function,
\begin{equation}
\bar\delta(\beta) = \frac{(2-\beta)}{{\int d^2x}} 
                    \frac{\int d^2x |(\triangle + k_0^2) v({\bf x})|^{\beta}}
                         {k_0^{2\beta} \langle |v({\bf x})|\rangle^{\beta}},
\label{defn}
\end{equation}
where $\langle|v({\bf x})|\rangle$ denotes the mean of $|v({\bf x})|$, 
and $\bar\delta(\beta)$ ( $0\le \beta < 2$) has been normalized to be 
scale invariant~\cite{gunAhof}. $\bar\delta(\beta)$ describes
configuration-independent aspects of textures and their
formation, aspects that are independent under repetition of the
experiment. It can be used to study multiple aspects of patterns
just as generalized dimensions~\cite{henApro} and singularity
spectra~\cite{halAjen} can be used to describe multiple aspects of
strange attractors.  In experiments presented in this paper, we
characterize the time evolution of patterns in a vertically oscillated
granular layer using $\bar\delta (\beta)$.

Before describing our experiments, we review numerical and analytical
work on the coarsening of patterns after a quench from an initial
featureless (or noisy) state. Most studies have focused on solutions
$u({\bf r},t)$ to the Swift-Hohenberg equation~\cite{croAhoh},
\begin{equation}
\label{sh}
 \frac{\partial u}{\partial t} = \bigl(\epsilon-(\triangle+k_0^2)^2
 \bigr)u-u^3 - \nu (\nabla u)^2 + \eta({\bf r},t), \end{equation}
 where $u({\bf r},t)$ is a two-dimensional scalar field, $\epsilon$ is
 the distance from pattern onset, $\nu$ is the strength of a
 non-variational term~\cite{gunArat}, and $\eta$ a random forcing term
 such that $\langle \eta({\bf r},t) \eta({\bf r'},t') \rangle = 2 F
 \delta({\bf r}-{\bf r'})\delta(t-t')$, where $F$ controls the
 strength of the noise.

For the formation of textures in (\ref{sh}), the width of the structure
factor $S(t)$ (i.e., the width of the peak in the azimuthal average of
$\langle\tilde u({\bf k},t) \tilde u(-{\bf k},t)\rangle)$, has been
shown to decay in two distinct stages~\cite{schAall}. $S(t) \sim
t^{-\frac{1}{2}}$ is obeyed until the peak amplitude of the field
$u({\bf x}, t)$ saturates, beyond which time the pattern coarsens and the decay becomes
slower. For $\epsilon=0.25$ and $\nu=0$, Cross and
Meiron~\cite{croAmei}, Elder et al.~\cite{eldAvin}, and Hou et
al.~\cite{houAsas} found that in this second region $S(t)$ decreased
as $t^{-\frac{1}{5}}$ when $F=0$, and as $t^{-\frac{1}{4}}$ when $F\ne
0$~\cite{eldAvin,houAsas}. Schober et al.~\cite{schAall} found that
for $F=0$ and $\nu=0$, $S(t) \sim t^{-\frac{1}{4}}$; the discrepancy
with earlier results could be due to the one-dimensionality of
their model.

The characterization of pattern formation in model systems
using $\bar\delta (\beta)$ has also shown the presence of two
stages: the emergence of domains characterized by $\bar\delta (\beta) \sim
t^{-\sigma_E(\beta)}$ with $\sigma_E(\beta) = \frac{1}{2} \beta$, and a slower
coarsening behavior~\cite{gunArat}. The relaxation exponent during coarsening
depends on the value of $\nu$ in (\ref{sh}), and is thus
expected to be system and model dependent~\cite{othermeasures}. 
We report analogous behavior in the present laboratory study of pattern
formation in an oscillated granular layer and present additional
differences between the two stages. 

Our experiments generate patterns in a layer of 0.165 mm bronze
spheres contained in a vertically oscillated circular container with
diameter 14 cm~\cite{melAumb}. The layer is four particle diameters
deep, and the cell is evacuated to 4 Pa to avoid any hydrodynamic
interaction between the grains and surrounding gas.  The control
parameters are the frequency $f$ of the sinusoidal oscillations and
the peak acceleration of the container, $\Gamma=(2 \pi f)^2 A^2/g$,
where $A$ is the maximum amplitude of the oscillation and $g$ is the
gravitational acceleration. As $f$ and $\Gamma$ are varied, a variety
of textures including striped, square or hexagonal
planforms are observed~\cite{melAumb}. Our analysis reported here is
restricted to patterns with square planforms. In the region of the
phase diagram studied, square patterns appear for increasing control
parameter at $\Gamma \approx 2.75$; the bifurcation is
subcritical. The geometry of the circular container allows relaxation
to an almost perfect square array through wavelength adjustment of the
pattern at the container wall over a distance of less than one
wavelength~\cite{squarecell}.

The granular surface is illuminated with a ring of LEDs
surrounding the cell. The light is incident at low angles and the
scattering intensity is a nonlinear function of the height of the
layer; scattering from peaks (valleys) creates bright (dark)
regions. The images are collected at the driving frequency and the
acceleration of the container is monitored during each run. $\Gamma$
is suddenly increased from its initial value of 2.2, where no
discernible structure is observed. As the grains are not in contact
with the container for part of their motion, we assume that the
initiation of the quench occurs at the first layer-plate collision
after the change of control parameter. The uncertainty in the time
origin is the dominant source of error in our measurements.

The top row of Fig. 1 shows that local square domains emerge and
coarsen to a final, almost perfect, square array. The bottom row shows
this process in Fourier space. A repetition of the experiment would
lead to similar, but not identical, intermediate states. Our aim is to
study configuration-independent aspects of this relaxation, and to
analyze their dependence on the control parameters $f$ and $\Gamma$.

Patterns such as those shown in the top row of Fig. 1 can be
represented by a discrete sampling of a smooth scalar field $v({\bf
x})$. The values of the field are known on a (typically square) grid,
but the analytical form of the field is unknown. The ingredients used
to deduce the form of the disorder function in (\ref{defn}) are its
invariance under arbitrary rigid motions of the texture and the nature
of the local planform. Local deviations of a pattern from squares (due
to curvature of the contour lines \cite{gunAhof}) contribute to
$\bar\delta(\beta)$ through the Laplacian, while variations of the
size of squares contribute via the choice of a ``global" $k_0$, which
is obtained from the field $v({\bf x})$ by minimizing the value of
$\bar\delta(1)$~\cite{gunArat}.  Unlike the information contained in
the structure factor, $|(\triangle + k_0^2) v({\bf x})|$ is a local
density of irregularities in the texture, and hence distinct
``moments" $\beta$ can be used to quantify multiple aspects of the disorder
density.

The images shown in Fig. 1 have sharp changes at the edges which lead
to high frequency contributions in their Fourier spectra. Their
removal through simple filtering causes contamination of the pattern
near the edges and leads to error in calculating
$\bar\delta(\beta)$. We use a method of noise filtering that involves
extending the image to a periodic one using ``Distributed
Approximating Functionals" (DAFs)~\cite{hofAgun,dafref}. Fourier
filtering can then be used on the extended image to eliminate high
frequency noise and undesirable harmonics~\cite{trends}. The evaluation of
$\bar\delta(\beta)$ requires an accurate estimation of $\triangle
v({\bf x})$, which typically amplifies any noise present in the
discrete, digital experimental data. A method for this calculation has
been presented in ~\cite{gunAhof}.

The behavior of $\bar\delta(1)$ for the relaxation of Fig. 1 is
shown by the symbols $\circ$ in Fig. 2.  The initial formation of
local rectangular domains and the final coarsening correspond to
distinct power law decays of $\bar\delta(1)$.  The transition
coincides with the saturation of the peak amplitude; i.e.,
nonlinear effects are negligible during domain formation (typically
4-5 oscillations) and become relevant during
coarsening~\cite{schAall}.

During the initial stage of pattern formation 
$\bar\delta(1) \sim t^{-0.49\pm0.02}$.
This dynamical scaling is similar to that describing the decay of the width of the
structure factor~\cite{eldAvin} and is related to the rate of domain growth in
phase ordering kinetics~\cite{bray,limitations}.  

Since nonlinear effects are negligible during domain formation, the
evolution can be modeled by (\ref{sh}) with the removal of the nonlinear and stochastic terms.
Numerical integration starting from states consisting
of random or Gaussian noise shows that 
$\int d^2x |u({\bf r},t)| \sim e^{\epsilon t} t^{-\frac{1}{4}}$ and 
$\int d^2x |(\triangle + k_0^2) u({\bf r},t)| \sim
e^{\epsilon t} t^{-\frac{3}{4}}$; consequently $\bar\delta(1) \sim
t^{-\frac{1}{2}}$. It has been shown analytically that the width of the structure factor 
for evolution of $u({\bf r},t)$ decays like $S(t) \sim
t^{-\frac{1}{2}}$ until the peak amplitude saturates~\cite{schAall},
providing further evidence for the interpretation that the spatiotemporal
dynamics is linear during the first stage.

Furthermore, during the domain forming stage, moments of the disorder
function decay as $\bar\delta(\beta) \sim t^{-\sigma_E(\beta)}$, where
$\sigma_E(\beta) \approx \frac{1}{2}\beta$ (see Fig. 3(a)). Analogous
behavior can also be seen by numerical integration of the linear terms
in (\ref{sh}). The linearity of $\sigma_E(\beta)$ and the value of the
proportionality constant suggest that multiple aspects of textures,
such as structure factor, curvature of contours and defect densities
decay via a single mechanism during the domain forming stage.

The initiation of domain coarsening coincides with the saturation of
peak heights of the granular layer (see Fig. 2); i.e., the latter
stages of pattern formation correspond to nonlinear spatiotemporal
dynamics of the field~\cite{schAall}. The observed scaling of the
disorder function is more complex.  For the evolution shown in Fig. 2
at $\Gamma=2.8$, $\bar\delta(1) \sim t^{-0.18}$ \cite{nonuniv}. This
is close, but not identical to the rate of relaxation of the structure
factor~\cite{eldAvin,croAmei,schAall}. Even though the moments
$\bar\delta(\beta)$ decay (approximately) as power laws
$t^{-\sigma_L(\beta)}$, the exponent is not linearly related to
$\beta$ as during the early phase, see Fig. 3(b).  There is a tendency
towards slower relaxation in Fig. 3(b) (compared to Fig. 3(a)) for
larger values of $\beta$. Since large values of $\beta$ preferentially
weight pattern defects, this is consistent with the slower relaxation
of the density of defects~\cite{houAsas} than that of the structure
factor~\cite{eldAvin,croAmei,schAall}.

The nonlinearity of $\sigma_L (\beta)$ implies that pattern relaxation occurs
on more than one length scale. If the ``envelope" of the texture depends on a
single length scale $L(t)$, the field can be locally expanded as
$u({\bf x},t) = Re \bigl[e^{i{\bf k\cdot x}} A({\bf X},t)\bigr]$, where ${\bf
X} = {\bf x}/L(t)$; thus
\vskip .15in
\centerline{$(\triangle + k_0^2) u({\bf x},t) = Re \Bigl[ e^{i{\bf k\cdot x}} \bigl(
\frac{2i}{L} {\bf k\cdot \nabla_X} A - \frac{1}{L^2} \triangle_X A \bigr)
\Bigr]$.}
\vspace{.15in}\noindent
Since $L(t) \gg k_0^{-1}$ during domain coarsening, the
last term can be neglected, leading to
\vskip .15in 
\centerline{$|(\triangle + k_0^2) u({\bf x}, t)| \approx \frac{2}{L(t)} {\bf k\cdot \nabla}
A \sim \frac{|u|}{L(t)}$.}
\vskip .15in\noindent
Consequently $\bar\delta(\beta) \sim 1/L^{\beta}(t) \sim
\bigl(\bar\delta(1)\bigr)^{\beta}$. Thus the nonlinearity of $\sigma_L(\beta)$
implies that the relaxation during coarsening occurs on multiple time scales.

Next, we briefly consider changes in the behavior of the disorder
function as the experimental system is driven further away from the
onset of patterns. The decay of $\bar\delta(\beta)$ during domain
formation remains unchanged, but the decay rate in the second region
decreases with increasing $\Gamma$. Similar behavior has
been observed with the addition of $\nu$ in (\ref{sh})~\cite{gunArat}.

We have shown that the formation of texture in a vertically oscillated
granular layer occurs in two distinct stages. During the initial stage
the spatiotemporal dynamics is essentially linear and the disorder
function obeys a universal power law $\bar\delta(\beta) \sim
t^{-\sigma_{E}(\beta)}$, with $\sigma_E(\beta)\simeq
\frac{1}{2}\beta$; this simple behavior is also observed in the
linearized Swift-Hohenberg equation. Nonlinearity of spatiotemporal
dynamics becomes relevant during domain coarsening and
$\bar\delta(\beta) \sim t^{-\sigma_{L}(\beta)}$ where
$\sigma_L(\beta)$ is a nonlinear function of $\beta$. The exponent
$\sigma_L(\beta)$ is model and parameter
dependent~\cite{gunArat}. Such non-universal, configuration
independent characteristics of pattern formation can be used to
determine the validity and limitations of model
systems~\cite{venAott}.

Although dynamical scaling has been reported in the formation of
patterns in model systems~\cite{eldAvin,croAmei,houAsas,gunArat}, ours
is the first reported observation of stages exhibiting trivial (i.e.,
a single scaling index such that $\bar\delta(\beta) \sim
\bigl(\bar\delta(1)\bigr)^{\beta}$) and nontrivial scaling during
pattern formation in an experimental system. The methods introduced
here are expected to have applications in studying other aspects of
textures, such as the quantitative description of patterns in magnetic
bubble material~\cite{molAgou,west}.

We have benefited from discussions with M. Golubitsky, J. B. Swift and
P. Umbanhower. The research at the University of Texas was supported
by the Engineering Research Program of the Office of Basic Energy
Sciences of the U.S. Department of Energy. Additional support came
from the Office of Naval Research (GHG), the Ames Laboratory of the
Department of Energy (DKH), the National Science Foundation (GHG,
DJK), and the R. A. Welch Foundation (DJK).

\begin{figure}
\epsfxsize=\textwidth
\centerline{\epsffile{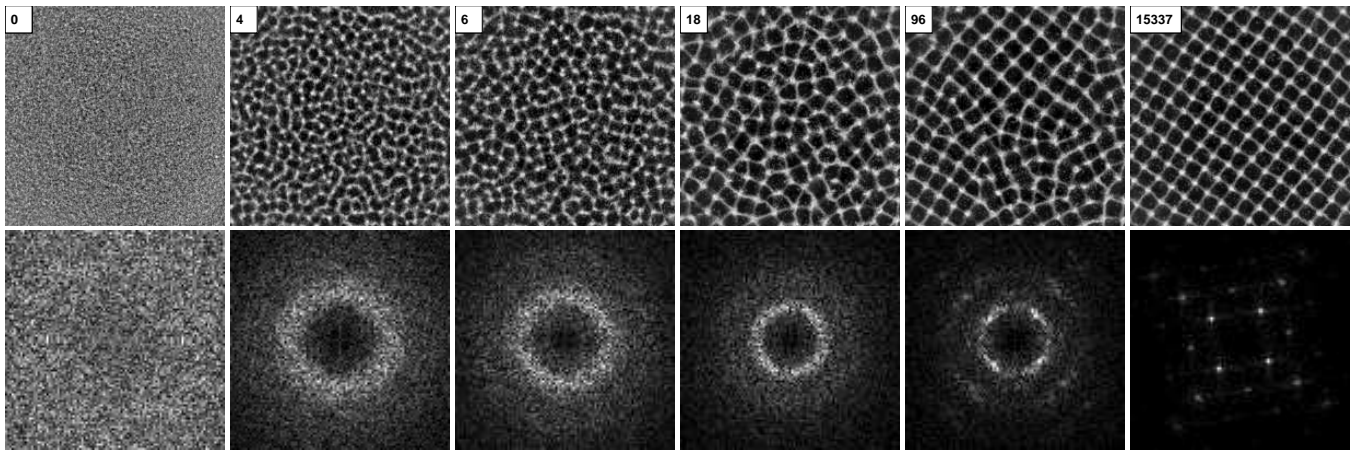}}
\vskip .25truein
\caption{Snapshots showing the emergence of a square spatial pattern
in a granular layer at $f = 27$ Hz and $\Gamma = 2.8$; the times given
in the upper left corner of each image are in units of container
oscillation periods. Each image in the first row is of the central 8
cm of the 14 cm diameter circular container. Each image in the second
row is the Fourier transform of the image above it. In the first row,
the first three frames show the emergence of local domains from a
uniform background, and the last three show the slower coarsening of
these domains to an almost perfect square array.}
\label{snapshots}
\end{figure}

\begin{figure}
\epsfxsize=3.5truein
\epsffile{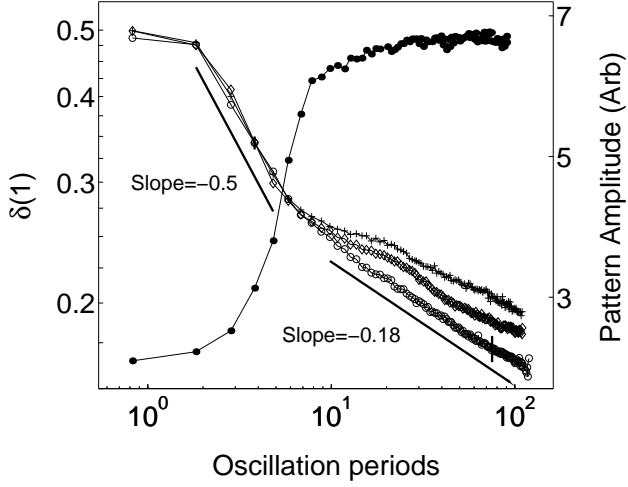}
\vskip 0.1in
\caption{The time evolution of the disorder function $\bar\delta(1)$
for square patterns at three different final container accelerations,
$\Gamma=2.8$ ($\circ$), $\Gamma=3.0$ ($\Diamond$), and $\Gamma=3.2$
($+$). Also shown is growth of the pattern amplitude at $\Gamma = 2.8$
($\bullet$). Rapid early growth in the domain-forming phase is
followed by a saturation of the amplitude in the coarsening phase.
Each curve is an average of 10 runs at the same control
parameters. The error bar at late times for $\circ$ shows typical
variation between distinct runs. The error bars at early times are the
size of the symbols.}
\label{relaxation} 
\end{figure} 

\vskip 2in

\begin{figure}
\hskip .3truein
\epsfxsize=2.5truein
\epsffile{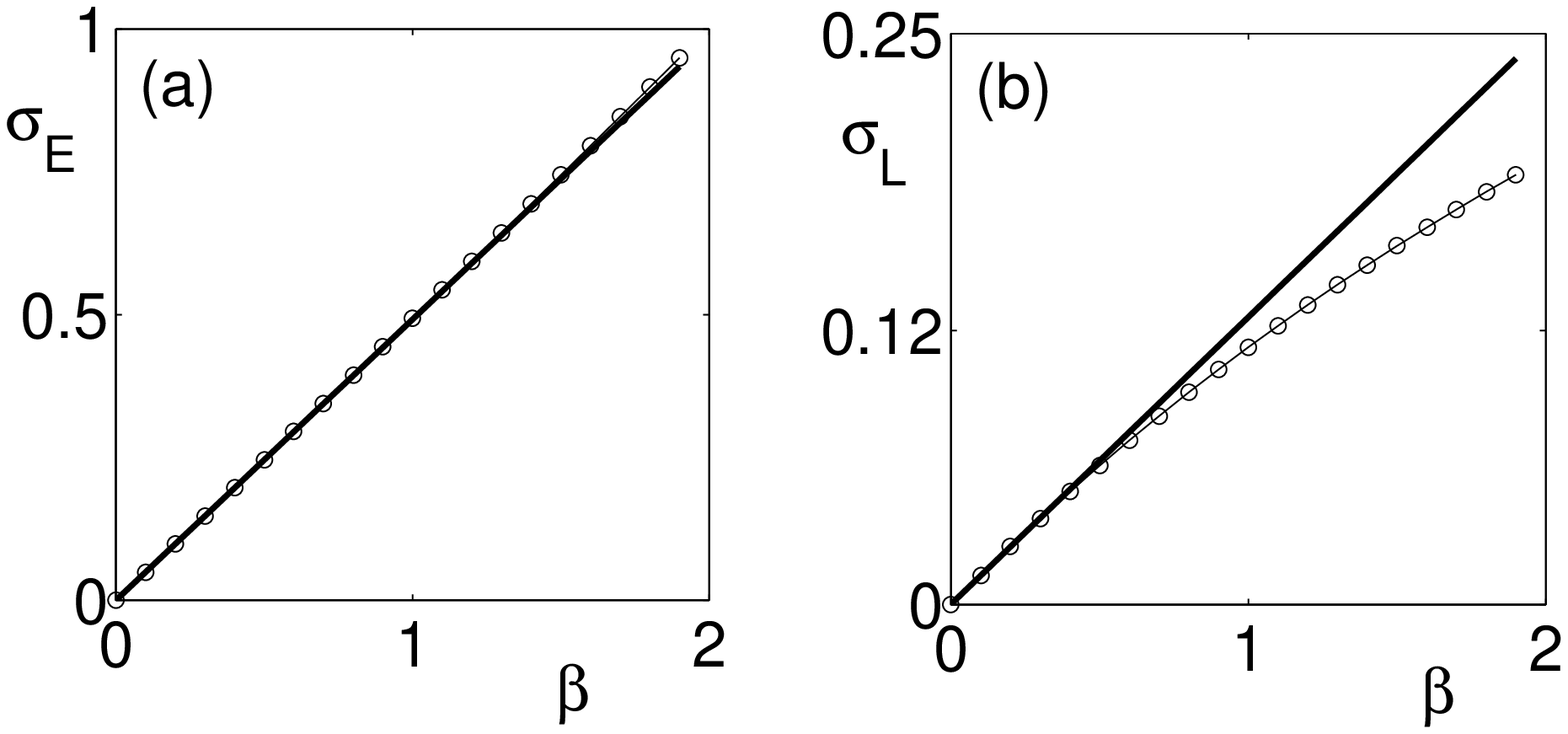}
\caption{The slopes $\sigma_E(\beta)$ \and $\sigma_L(\beta)$ of the
curves $\bar\delta(\beta)$ during (a) domain formation and (b) coarsening. The
results are for a single run at control parameters $f=27$ Hz and
$\Gamma=3.0$.  The bold straight lines are drawn to guide the eye. In
(a), the bold line has slope $=1/2$ and in (b), the bold line has slope $=0.12$.}
\label{slopes}
\end{figure}


\begin{thebibliography}{99}
\bibitem{othermeasures} Additional quantities 
  have been measured in the second relaxation region. Hou et
  al. measured total domain wall length in $v$ as a function of
  time. For $F=0$, it was shown to decay as $t^{-\frac{1}{4}}$,
  while for $F \neq 0$ the decay was proportional to $t^{-0.3}$. Cross et
  al. showed that a stripe orientation correlation field 
  decayed as $t^{-0.24}$. The interpretation of these
  results is that different measures probe different features of the
  pattern, and these features evolve via different mechanisms.
\bibitem{squarecell} Experiments done in square containers also
exhibit our main results. However, the formation of a
single domain takes a much longer time, as the final pattern
orientation is strongly constrained by the boundaries.
\bibitem{trends} Calculations reported here are performed by a
preliminary estimate of the wave-vector $k_t$ (the wave-vector of the
peak of the radial average of the power spectrum), followed by the
filtering of spectral components outside $[0.7k_t, 1.3k_t]$. The
trends reported here do not depend on the range of spectral components
retained.
\bibitem{nonuniv} This index is not expected to be universal; see Ref. [9].
\end{thebibliography}
\end{document}